\documentclass{jaa}
\usepackage{natbib}
\usepackage{hyperref}
\usepackage{graphicx}
\usepackage{aasmacros}
\usepackage{color}
\usepackage{booktabs} % for nicer tables

\hypersetup{
    colorlinks=true,
    linkcolor=cyan,
    filecolor=magenta,      
    urlcolor=red,
    citecolor=blue,
    pdftitle={K. Sasikumar Raja},
    }
\usepackage{graphicx}

\graphicspath{{./}{figures/}}
\DeclareUnicodeCharacter{202F}{\,} % or \, for thin space, ~ for no line break

\begin{document}\sloppy

%%paper title
%%For line breaks \\ can be used within title
\title{%Insights into Coronal Turbulence from Aditya-L1/VELC 5303 \AA\ Emission %Line Observations 
Aditya-L1/VELC observations of CME associated broadening
of 5303{\AA} coronal emission line}

%%author names are separated by comma (,)
%%use \and before the last author name
%%use a * along with the number separated by comma
%% for the  author for correspondence
%%\textsuperscript{number} is used for affiliation
%%\affilOne, \affilTwo etc., upto \affilTwentyfive is possible
%%Please note the first letter after \affil is capitalised in the command
%%

\author{K. Sasikumar Raja\textsuperscript{1*},
V. Muthu Priyal\textsuperscript{1},
R. Ramesh\textsuperscript{1},
Jagdev Singh\textsuperscript{1},
Prasad Subramanian\textsuperscript{2}
%P. Savarimuthu \textsuperscript{1},
%Nagashree S \textsuperscript{1} and
%Yuvashree E \textsuperscript{1}
}

\affilOne{\textsuperscript{1} Indian Institute of Astrophysics, II Block, Koramangala, Bengaluru - 560 034, India.\\}

\affilTwo{\textsuperscript{2} Indian Institute of Science Education and Research, Pashan, Pune - 411 008, India\\}

%%escape two column mode for title, affiliation and abstract
%%by giving \twocolumn command as shown
\twocolumn[{
	\maketitle
	%%include \corres to print the corresponding author Email id
	\corres{sasikumar.raja@iiap.res.in}
	%%include \msinfo for manuscript information such as received, revised and accepted dates
	\msinfo{1 January 2015}{1 January 2015}
	
	%%abstract
	\begin{abstract}
		
	We present Aditya-L1/VELC spectroscopic observations of 5303{\AA} coronal emission line widths before and after coronal mass ejections (CMEs) which showed coronal dimming. The sit \& stare mode of observations enabled us to study the changes in the line widths for the `limb' CMEs noticed on 16 July 2024 and 05 August 2024. The emission line widths during the pre-CME phase are higher than the thermal values in both the cases. After the onset of the CMEs, the widths increased further by   
	${\approx}15\%$ on 16 July 2024 and ${\approx}7\%$ on 05 August 2024. 
	We find that the power spectral density (PSD) distributions of the line widths for the two events exhibits a power-law behavior.  The PSD slopes, measured before and after the CMEs are nearly the same, and close to the Kolmogorov slope of ${-5/3}$.  The results suggest that the observed larger than thermal width of the 5303{\AA} emission line, before and after CMEs, is mostly due to turbulence. The 
	additional increase in the line widths after the onset of the CMEs in both the cases are likely because of enhanced turbulence caused by the CME associated coronal dimmings and subsequent coronal magnetic field reconfiguration.
	\end{abstract}
	
	%%insert keywords separated by 3 hyphens using \keywords{words}
	\keywords{Sun---Corona---Coronal emission line---Coronal mass ejections---Turbulence}
}]

%%close the twocolumn escape here

%%include \doinum{number} for the DOI number in the header
%%include \volnum{number} for the volume number in the header
%%include \year{yyyy} for  year of publication in the header
%%include \pgrange{num--num} page range of article in the header
%%include \artcitid{num} for the article citation id
%%include \lp to print last page of the article
%%include \setcounter{page}{pagenum} for the exact starting page of the article

\doinum{12.3456/s78910-011-012-3}
\artcitid{\#\#\#\#}
\volnum{000}
\year{0000}
\pgrange{1--}
\setcounter{page}{1}
\lp{12}

\section{Introduction}
Aditya-L1 is the first Indian space mission dedicated to exploring the Sun and the solar atmosphere from the Sun-Earth Lagrangian L1 point \citep{Parate2025}. It was launched on 2 September 2023 from Sriharikota, India. The mission comprises four remote sensing payloads, operating across a wide frequency range from near-infrared to hard X-rays, and three in-situ payloads. After 126 days of travel, Aditya-L1 was inserted into a halo orbit around the L1 point on 6 January 2024.

The Visible Emission Line Coronagraph (VELC), an internally occulted coronagraph, is the prime payload of Aditya-L1 \citep{BRP2017,  BRP2023, RK2018, JS2013, Singh2025}. VELC has four channels: one imaging channel to observe the corona in the heliocentric distance ($r$) range 1.05\,-\,3$R_{\odot}$, and three spectroscopic channels that acquire spectra at wavelengths of 5303{\AA}, 7892{\AA}, and 10747{\AA} in the range $r$\,=\,1.05\,-\,1.5$R_{\odot}$. 
The 10747{\AA} spectroscopic channel can also be used for spectropolarimetric observations \citep{KSR2022, VSN2022a, VSN2024}.
The performance of VELC was evaluated at the Prof. MGK Menon Laboratory at CREST campus of the Indian Institute of Astrophysics (IIA). \citet{BRP2023, Mishra2024, VSN2024} have reported the related results. The in orbit performance of VELC can be inferred from \citet{Singh2025}.

In this article, we discuss the observations obtained using the 5303{\AA} (also called the green line) spectroscopy channel. Green line observations by VELC, covering the range 1.05\,-\,1.5$R_{\odot}$, are unique and offer valuable insights into the coronal structures \citep{Ramesh2024,MP2025ApJ...983..171M}. VELC green line observations can be carried out in two modes:
(i) Sit \& stare mode, where the slit is fixed at a user defined specific location to observe temporal evolution of coronal structures such as loops, prominences, streamers, and CMEs;
(ii) Raster scan mode, where the field of view (FoV) from -1.5$R_{\odot}$ to +1.5$R_{\odot}$ is scanned to acquire spectral images that are subsequently used to reconstruct images of the solar corona.

Using data from these two modes, particularly the sit \& stare observations, it is possible to measure the characteristic parameters of the emission line like the intensity, width, and Doppler velocity in different coronal regions and structures.

A recent study by \citet{Ramesh2024} reported a coronal dimming event observed in the 5303{\AA} channel using the sit \& stare mode. The authors found that the emission line width increased by approximately 15\% after the CME, suggesting that the enhancement was due to increased turbulence.  
Spectroscopic detection of turbulence in post-CME current sheets in the range 1.5\,-\,1.7$R_{\odot}$ using Ultraviolet Coronagraph Spectrometer (UVCS) onboard SoHO were earlier reported by \citet{Bemporad2008}.  A recent study by \citet{Shen2023} indicates that the turbulent bulk plasma flows in the current sheets and
flare loop-top regions are responsible for the non-thermal broadening of the 
Fe\,XXI emission line observed with the Interface Region Imaging Spectrograph (IRIS) spacecraft. Studies on Alfv{\' e}nic turbulence using data obtained with the ground-based Coronal Multichannel Polarimeter (UCoMP) can be found in \citet{Moortel2014,Sharma2023}. The slope of the Power Spectral Density (PSD) distribution is an useful parameter to infer whether the observed nonthermal broadening of a coronal emission line can be due to turbulence. In this article, we investigate the PSD of the 
5303{\AA} coronal emission line widths before and after the CME using data obtained with the Aditya-L1/VELC.
% to better understand the role of turbulence. 
\section{Observations and data analysis}

\begin{figure*}[!ht]
	\centering
	\includegraphics[width=0.95\linewidth]{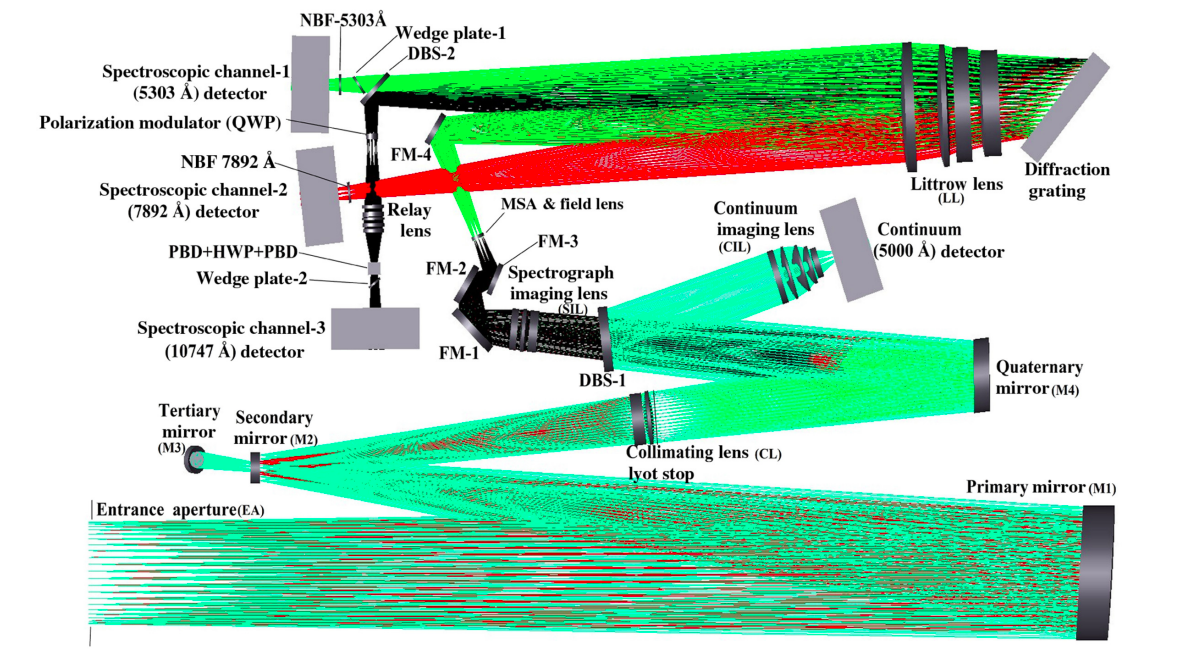}
	\caption{Optical layout of Visible Emission Line Coronagraph.}
	\label{fig:layout}
\end{figure*}
VELC is an internally occulted, reflection-type coronagraph designed to observe the solar corona as close as $r$\,=\,1.05$R_{\odot}$. Figure \ref{fig:layout} shows the optical layout of VELC.
Light entering through the Entrance Aperture (EA), which has a diameter of 147 mm, is reflected by the primary mirror (M1) with a clear aperture of 192 mm, and is then focused onto the secondary mirror (M2). Since M2 is positioned at the focal point of M1, an elliptical central hole (with major and minor diameters of 13.05 mm and 12.95 mm, respectively) in M2 allows the disk light and coronal light till
$r$\,=\,1.05$R_{\odot}$ to pass through. The above light beam is reflected/directed into the deep space by the tertiary mirror (M3). Only the coronal light in the distance range 

$r$\,=\,1.05\,-\,3.0$R_{\odot}$
is reflected toward the Quaternary mirror (M4) via the Collimator Lens Assembly (CLA) and the Lyot stop.

M4 then reflects the coronal light toward the Dichroic Beam Splitter-1 (DBS-1). DBS-1 reflects light with wavelengths below 5150{\AA} and transmits light with wavelengths above 5150{\AA}. The reflected light is imaged onto an sCMOS detector using the Continuum Imaging Lens Assembly (CIL). This imaging channel has a central wavelength of 5000{\AA} and a spectral bandwidth of 10{\AA}.

The transmitted light is focused by the Spectrograph Imaging Lens Assembly (SIL). The SIL, together with mirror M4, images the light onto the Multi-Slit Assembly (MSA) via fold mirrors FM1, FM2, and FM3. The MSA contains four slits, each with a width of 50${\mu}$m. This corresponds to 9.6$^{\prime\prime}$ in the plane of the sky, and 0.218{\AA} in the spectral image. FM1 and FM2 are mounted on a Linear Scan Mechanism (LSM). By moving the LSM, the coronal image projected onto the slit is shifted, enabling the spectrograph to observe different coronal regions. The LSM operates within a range of –1\,mm to +1\,mm, with step sizes in integral multiples of $10{\mu}$m. Using the LSM, it is possible to either fix the slit at a single location or scan across the FoV.

The light from the MSA is directed by FM4 towards the Littrow Lens Assembly (LL), which functions as both a collimator and camera lens. A diffraction grating serves as the dispersing element, and the LL focuses the dispersed beam onto various detectors. The reflected light from DBS-2 passes through a rotatable retarder (i.e., a Quarter Wave Plate), a relay lens assembly, and a PBD+HWP+PBD system (where PBD is a Polarized Beam Displacer and HWP is a Half-Wave Plate) before being focused onto the IR detector. This PBD+HWP+PBD configuration splits the input spectra into ordinary and extraordinary rays \citep{KSR2022, VSN2023a, VSN2024}.

\begin{figure*}[!ht]
    \centering
    \includegraphics[scale=0.11]{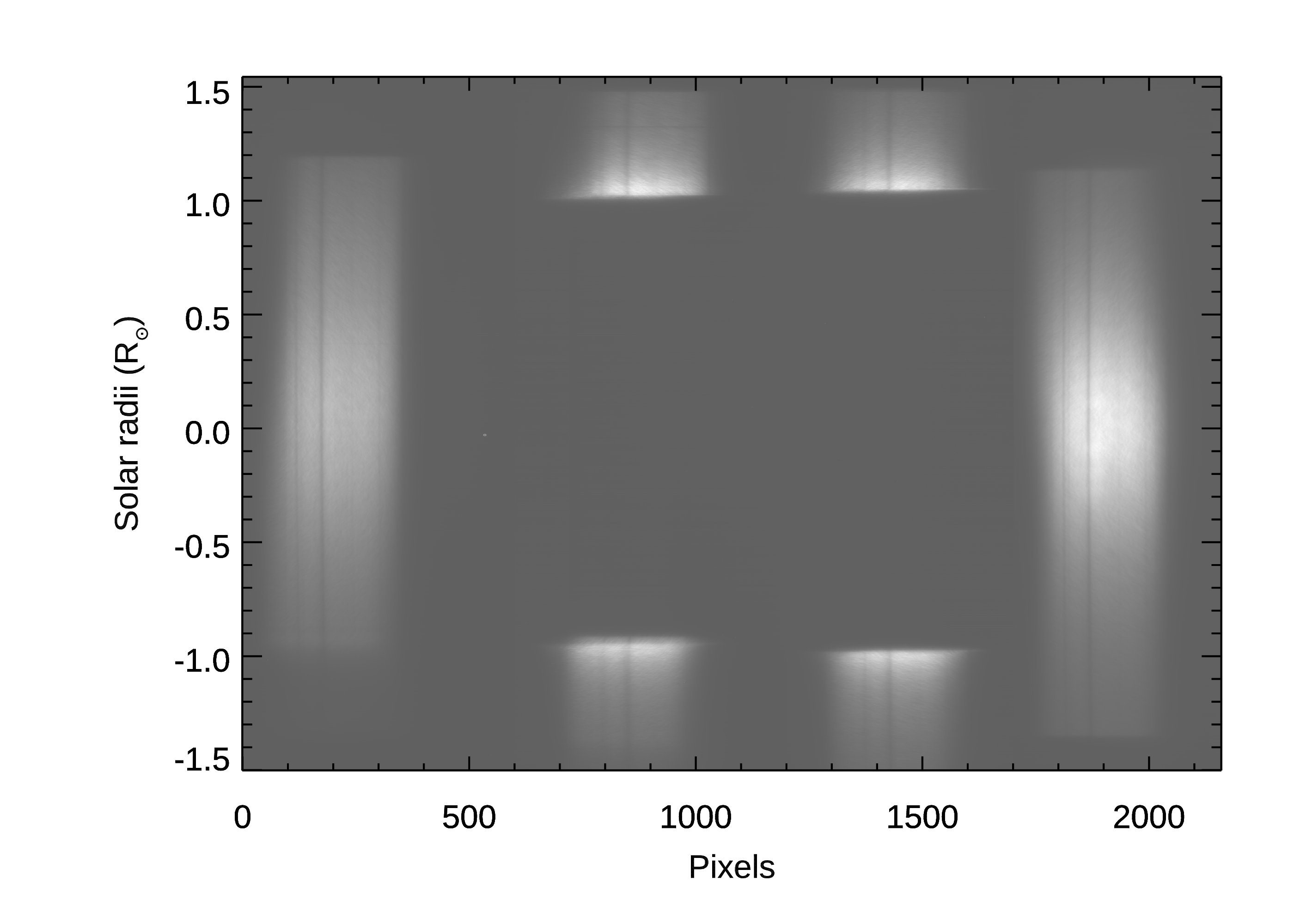}
    \caption{Sit \& stare mode of observations in VELC 5303{\AA} channel observed on 16 July 2024 during 12:30-22:30 UT. The four spectra corresponds to the four slits of VELC. Slit 1 is to the far left and slit 4 is to the far right. The gap in the middle for slits 2 \& slit 3 are due to the occulter. Since the FoV is circular, upper and lower parts of slits 1 \& 4 are not illuminated unlike slits 2 \& 3.}
    \label{fig:four_spectra}
\end{figure*}

In this article, we present sit \& stare mode observations carried out on 16 July 2024 and 5 August 2024 using the 5303{\AA} channel. On 16 July 2024, data were acquired in 2x1 spatial binning, low-gain (1x) mode with an exposure
time of 5\,sec. Slit 4 was positioned at 1.05$R_{\odot}$ on the West limb. Data was continuously acquired with a cadence of 51\,sec for a duration of 10\,hours (12:30\,-\,22:30 UT). The four vertical lines in Figure \ref{fig:four_spectra} indicate the four spectra corresponding to four spatial locations. The horizontal axis corresponds to the solar east-west direction, and the vertical axis corresponds to the solar north-south direction. Furthermore, the vertical axis represents the spatial direction of the slits, and the horizontal axis represents the spectral direction of the slits. The discontinuities in the spectra of slits 2 and 3 are due to the presence of the occulter. The detector has 2560\,${\times}$\,2160 pixels. The plate scale is 1.25$^{\prime\prime}$ per pixel. 

\begin{figure*}[!ht]
    \centering
    \includegraphics[scale=0.91]{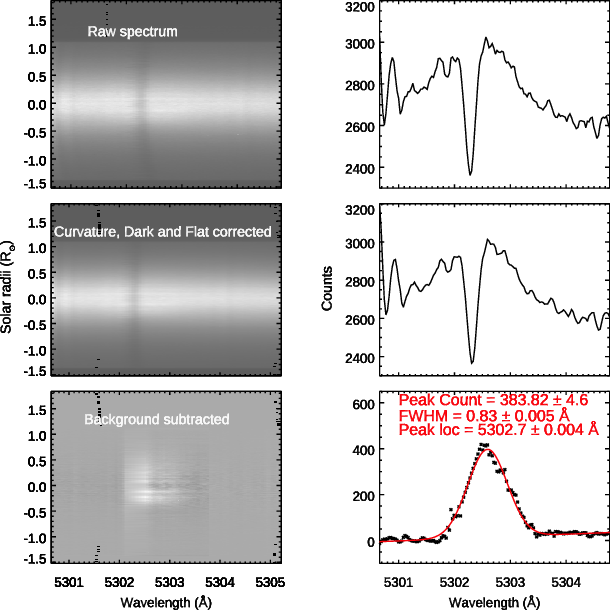}
    \caption{Left panel: The upper image is the instantaneous raw spectrum generated using data obtained with slit 4 on 16 July 2024. The exposure time is 5\,sec. Vertical axis is the spatial direction along the slit. The spectral dispersion is across the slit, along the horizontal axis. The middle image is the spectrum obtained after corrections for curvature, and detector response (dark current \& flat-fielding). 
Right panel: Spectral profiles corresponding to the centre of the slit in the respective left panel images. The red colour profile is the Gaussian fit to the observed emission line profile after the abovementioned corrections. The 5303{\AA} coronal spectrum, after subtracting the continuum background from the spectral profile at each spatial location along the slit, is shown in the lower left panel.}    
    \label{fig:dataprocedure}
\end{figure*}

For the entire spectral dataset, dark subtraction and flat-field correction are carried out. Following the curvature correction and background subtraction are performed. Once all corrections are completed, 
the emission profile (from each row in the detector corresponding to slit 4) is  extracted and the peak counts, line widths, and Doppler velocity at each spatial location along the slit are measured by fitting a Gaussian. The average wavelength of the emission line from large number of high spectral resolution observations with the 25\,cm Norikura coronagraph is 
5302.8{\AA}, which corresponds to 5304.3{\AA} in vacuum \citep{Singh2004,Singh2006}. This wavelength is used as reference to compute the Doppler velocity. We verified that the relative changes in the computed Doppler velocities are the same as the corresponding changes estimated using the wavelength of the dominant solar disk absorption line in the VELC data (5302.3{\AA} in air) as the reference. Figure \ref{fig:dataprocedure} shows an example of the procedure adopted to obtain the emission line and its parameters. More details can be found in \citet{Singh2025}.
%\citet{Ramesh2024, MP2025ApJ...983..171M}.
The instrumental width, inferred from observations of the above mentioned absorption line is ${\approx}$0.272{\AA}. 
It is used for correcting the observed line widths.
The spectral resolution per pixel, important for Doppler velocity measurements using the shifts in the pixel location of the maximum in the emission line, is 0.0284{\AA}. The equivalent velocity is 1.6\,km/s.  

\begin{figure*}[!ht]
    \centering
    \includegraphics[scale=1]{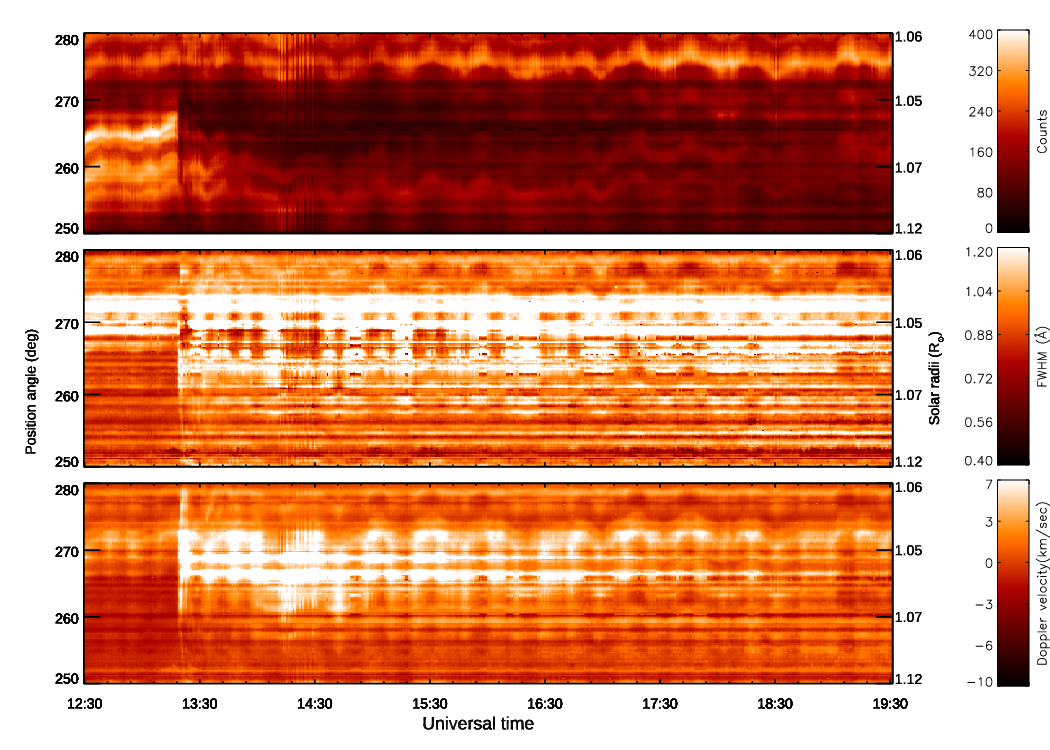}
    %16072024_3_in_1_Spectrum_plot_with_cb.eps}
 %   16072024_3_in_1_spectrum_plot.png}
       \caption{Sit \& stare mode of observations with VELC 5303{\AA} channel when slit 4 is placed at a heliocentric distance of 
    $\approx$1.05$R_{\odot}$ on the west limb. The upper panel shows the temporal variations in the peak emission line intensities of 
    coronal structures in the positional angle range 
    $250{^\circ}$\,-\,$280{^\circ}$. The corresponding heliocentric distances are shown in the right hand side y-axis. Since it is a straight slit, the heliocentric distances increase on either side of the slit center which is at $r$\,=\,1.05$R_{\odot}$ in this case. The corresponding position angle is $270^{\circ}$. Coronal dimming is seen during 13:18\,-\,19:00 UT in the line intensity shown in the upper panel (see, \citealp{Ramesh2024} for details). The middle and lower panels show the widths and Doppler velocities of the emission line profiles corresponding to the intensities in the upper panel.}   
    \label{fig:ss2d}
\end{figure*}

The procedure shown in Figure \ref{fig:dataprocedure} is repeated for all the observed spectra, and the results are compiled as a two-dimensional array. The image obtained from the latter is shown in Figure~\ref{fig:ss2d}. The center of slit 4 is kept at $r$\,=\,1.05$R_{\odot}$ using the LSM, for this observation. A similar analysis was carried out for the observations on 5 August 2024 during 01:15-22:15 UT. The center of slit 4 is $r$\,=\,1.12$R_{\odot}$ in this case. The detector configurations and mode of operations are tabulated in Table \ref{tab:tab1}.

\begin{figure*}[!ht]
    \centering
    \includegraphics[scale=0.3]{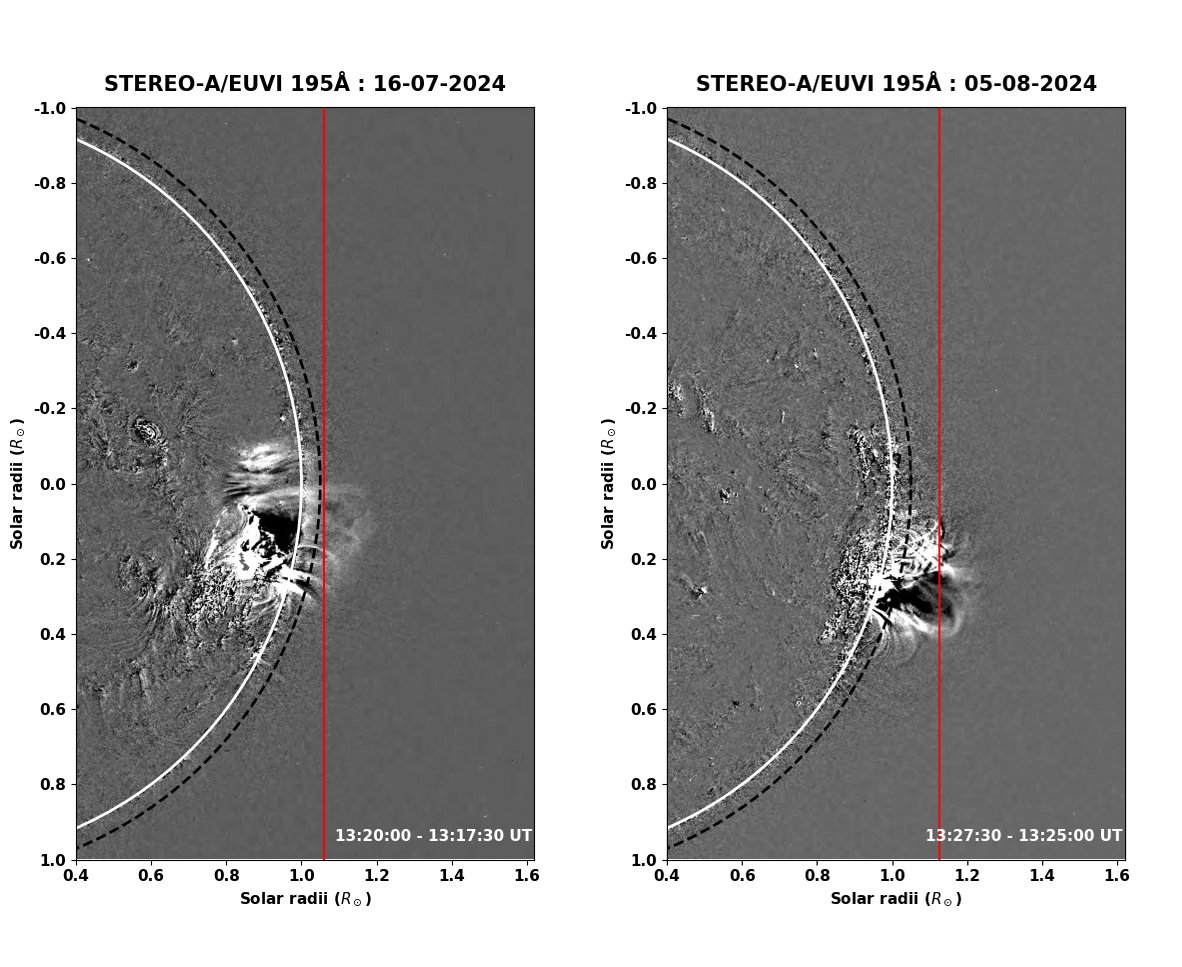}
       \caption{STEREO-A 195{\AA} difference images of the solar corona showing the coronal dimmings associated with the CMEs observed on 16 July 2024 (left panel) and 5 August 2024 (right panel). Solar north is straight up and east is to the left. The white and black semi-circles indicate the solar limb ($r$\,=\,1$R_{\odot}$) and the VELC occulter ($r$\,=\,1.05$R_{\odot}$). The vertical red line indicates the location of the straight slit 4 of VELC on the west limb.}
    \label{fig:context}
\end{figure*} 

\begin{table*}[h!]
	\centering
	\begin{tabular}{clcc}
		\toprule
		\multicolumn{2}{c}{} & \multicolumn{2}{c}{Value} \\
		\cmidrule(lr){3-4}
		S.No. & Parameter & 16 July 2024 & 5 August 2024 \\
		\midrule
		1 & Exposure time (sec) & 5 & 5 \\
		2 & Gain       & 1x & 1x \\
		3 & Observation mode & Snapshot & Snapshot \\
		4 & Spatial binning (pixels) & 2 & 2 \\
		5 & Spectral binning (pixels) & 1 & 1 \\
		6 & Cadence (sec)    & 51 & 51  \\
		7 & Observation period  & 12:30\,-\,22:30\,UT & 01:15\,-\,22:15\,UT \\
		8 & Observation method    & Sit \& Stare & Sit \& Stare \\	
		9 & Location of center of slit 4 & &  \\
		& on the west limb & 1.05$R_{\odot}$ & 1.12$R_{\odot}$ \\
		\bottomrule
	\end{tabular}
	\caption{Detector configurations and operation mode of VELC on 16 July 2024 and 5 August 2024}
	\label{tab:tab1}
\end{table*}

\section{Results and discussions}

\begin{figure*}[!ht]
    \centering
    \includegraphics[width=0.75\linewidth]{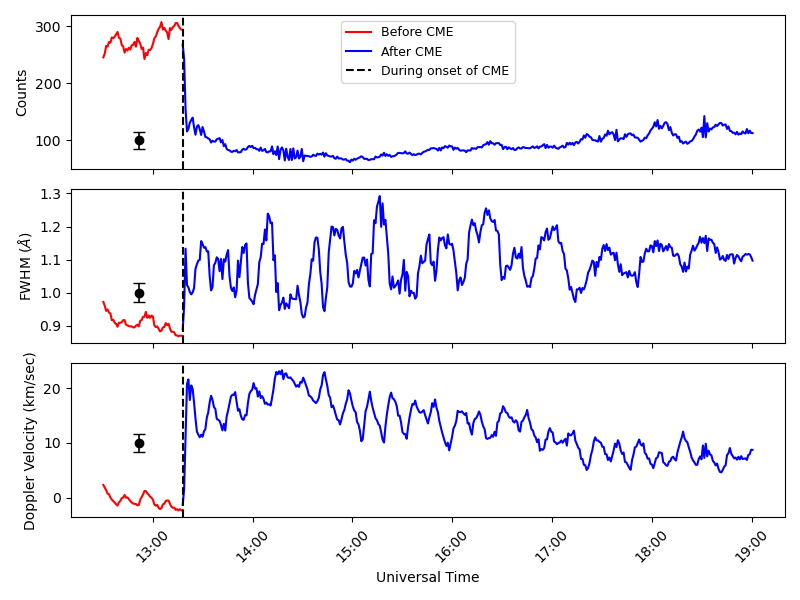}
    %{light_curve_16072025.png}
    \caption{The upper panel shows the temporal variation of the peak intensity of the emission line observed on 16 July 2024, averaged over the position angle range $260{^\circ}$\,-\,$270{^\circ}$ (see Figure \ref{fig:ss2d}). The middle and lower panels display the corresponding emission line width and Doppler velocity, also averaged over the same range. In both panels, the regions before and after the vertical dashed line represent the periods before and after the CME onset, respectively. The black colour error bars on the left side in each panel represent the measurement uncertainities.}
    \label{fig:light_curve_16072024}
\end{figure*}

The observations on 16 July 2024 and 05 August 2024 revealed CME associated coronal dimmings. Figure~\ref{fig:context} shows the context images and the location of the VELC slit 4. The peak intensities, emission line widths and Doppler velocities of the event observed on 16 July 2024, averaged over position angles $260^{\circ}$\,-\,$270^{\circ}$, are shown in Figure \ref{fig:light_curve_16072024}.
During this event, the mean line width increased to \(1.2 \pm 0.07\){\AA} after the CME onset, compared to \(0.91 \pm 0.02\){\AA} before the CME event. Similarly, the Doppler velocity increased from \(-0.63 \pm 1.08\) km/sec to \(12.94 \pm 4.75\) km/sec.

\begin{figure*}[!ht]
	\centering
	\includegraphics[width=0.7\linewidth]{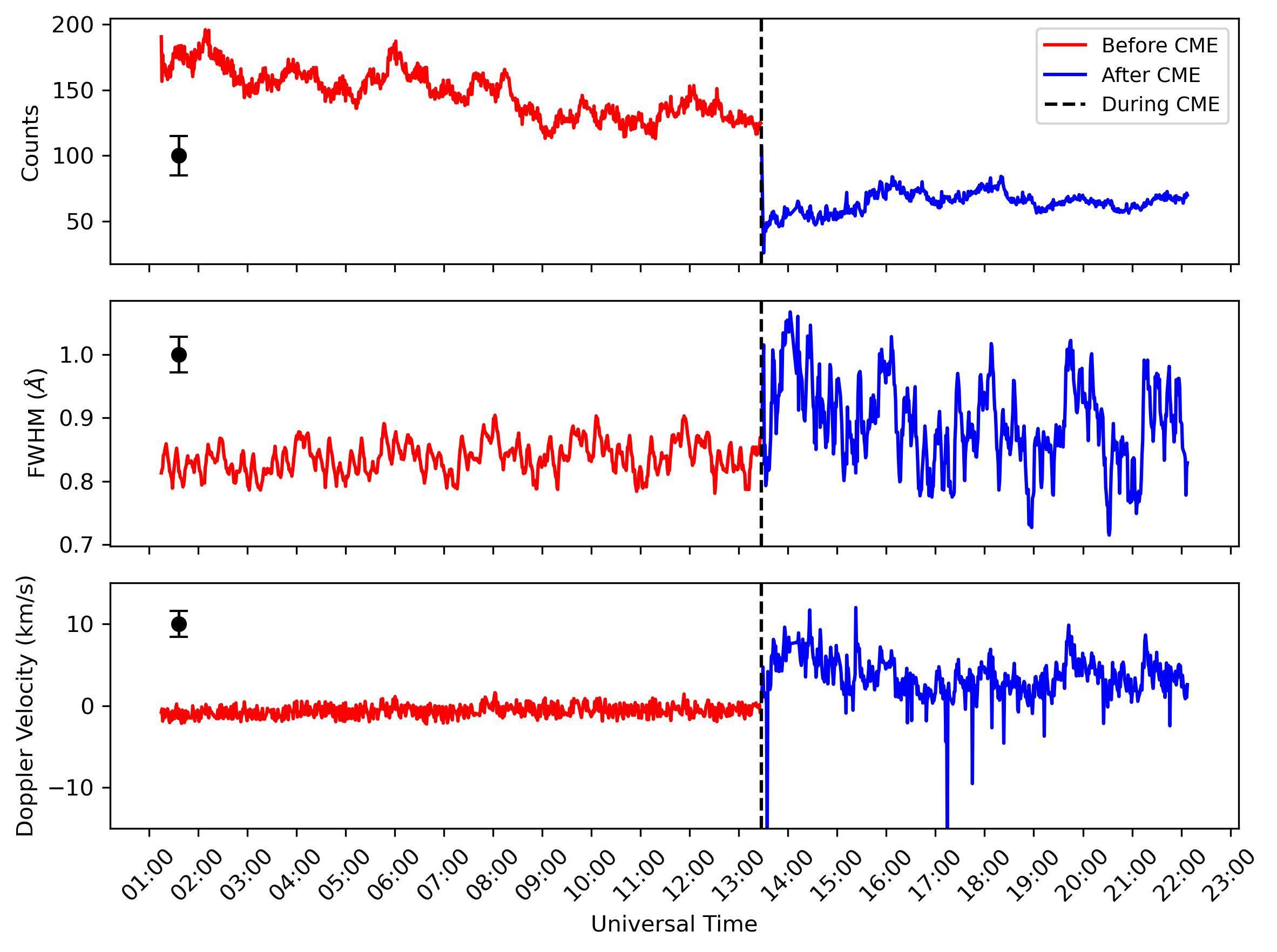}
	%{light_curve_05082025.png}
	\caption{The upper panel shows the temporal variation of the peak intensity of the emission line observed on 5 August 2024, averaged over the position angle range $240{^\circ}$\,-\,$260{^\circ}$. The middle and lower panels display the corresponding emission line width and Doppler velocity, also averaged over the same range. In both panels, the regions before and after the vertical dashed line indicate the periods before and after the CME onset, respectively. The black colour error bars on the left side in each panel represent the measurement uncertainities.}
	\label{fig:light_curve_05082024}
\end{figure*}

Similarly, the averaged peak intensities, line widths and Doppler velocities of the event observed on 05 August 2024, averaged over position angles $240^\circ - 260^\circ$, are shown in Figure \ref{fig:light_curve_05082024}. During this event, the mean line width increased to \(0.89 \pm 0.07\) after the CME onset, compared to \(0.84 \pm 0.03\) before the CME event. Similarly, the Doppler velocity increased from \(-0.65 \pm 0.67\) km/sec to \(3.49 \pm 2.56\) km/sec. The oscillatory pattern in the plots are due to spacecraft pointing issues \citep{Singh2025,MP2025ApJ...983..171M}, the cause for which is unknown. The dominant periodicities inferred from FFT analysis of both the data sets in the present case, prior to as well as after the onset of the CMEs, are 
${\approx}$13 \& 33\,min. Wavelet analysis indicate that these two periodicities lasted the entire observation duration. Since they are noticed in both 16 July 2024 and 05 August 2024 data, we assume them to be because of the pointing issue mentioned above. An inspection of the VELC observations during different epochs reveal that periodicities are always noticeable in the data.  

\citet{Ramesh2024} reported coronal dimming in the 5303{\AA} line for the CME event of 16 July 2024.
%, after 13:18 UT. 
The green emission line widths increased by 
%changed by 
approximately 15\% 
%before and 
after the CME. The broadening of the emission line widths was attributed to enhanced turbulence resulting from the restructuring of coronal magnetic field lines 
%and small-scale magnetic reconnections 
at the source region following the CME onset.
%A change in turbulence is believed to have caused the increase in the line width. Therefore, 
In this study, we investigated the turbulence spectra before and after the CME onset. Note that the association between CMEs and coronal turbulence in the near-Sun corona ($r{<}2R_{\odot}$) is known from radio observations \citep{Carley2021,Ramesh2023}. Observational signatures of coronal turbulence at large distances from the Sun are also well known \citep{Man2000, Ram2001,KSR2017}. CME observations at 1\,AU indicate that turbulent fluctuations can be used as an input for space weather prediction \citep{Deep2025}.

It is known that the slope of Power Spectral Density (PSD) distribution reveals how energy is distributed across different spatial or temporal scales. If the slope is close to -5/3, it corresponds to the inertial range, where energy cascades from larger to smaller scales without dissipation \citep{kolmogorov1941}; if the slope is steeper than -5/3, it indicates that energy is dissipated \citep{Bas94,Ing2015a,KSR2016,KSR2019,KSR2021}. Though the Kolmogorov slope (-5/3) describes spatial description of energy distribution in a turbulent medium, %there is connection between 
the temporal spectrum of the variations in the amplitudes measured at a fixed location in the medium can be considered to represent    
the spatial spectrum of the amplitude variations measured at different locations in the same medium, if the turbulence evolves slowly compared to the speed at which the mean flow carries them along \citep{Taylor1938,Perez2021}. 
In the present case, the amplitudes of the nonthermal widths (i.e. the equivalent nonthermal velocities) and Doppler velocities as well as their variations are smaller than the propagation speed of the CME plasma across the slit. So, the slope of the PSDs estimated using the sit \& stare observations indicates the spectral indices associated with the underlying spatial spectrum of turbulent fluctuations in the above mentioned line parameters. 

\begin{figure*}[!ht]
    \centering
    \includegraphics[width=.6\linewidth]
    {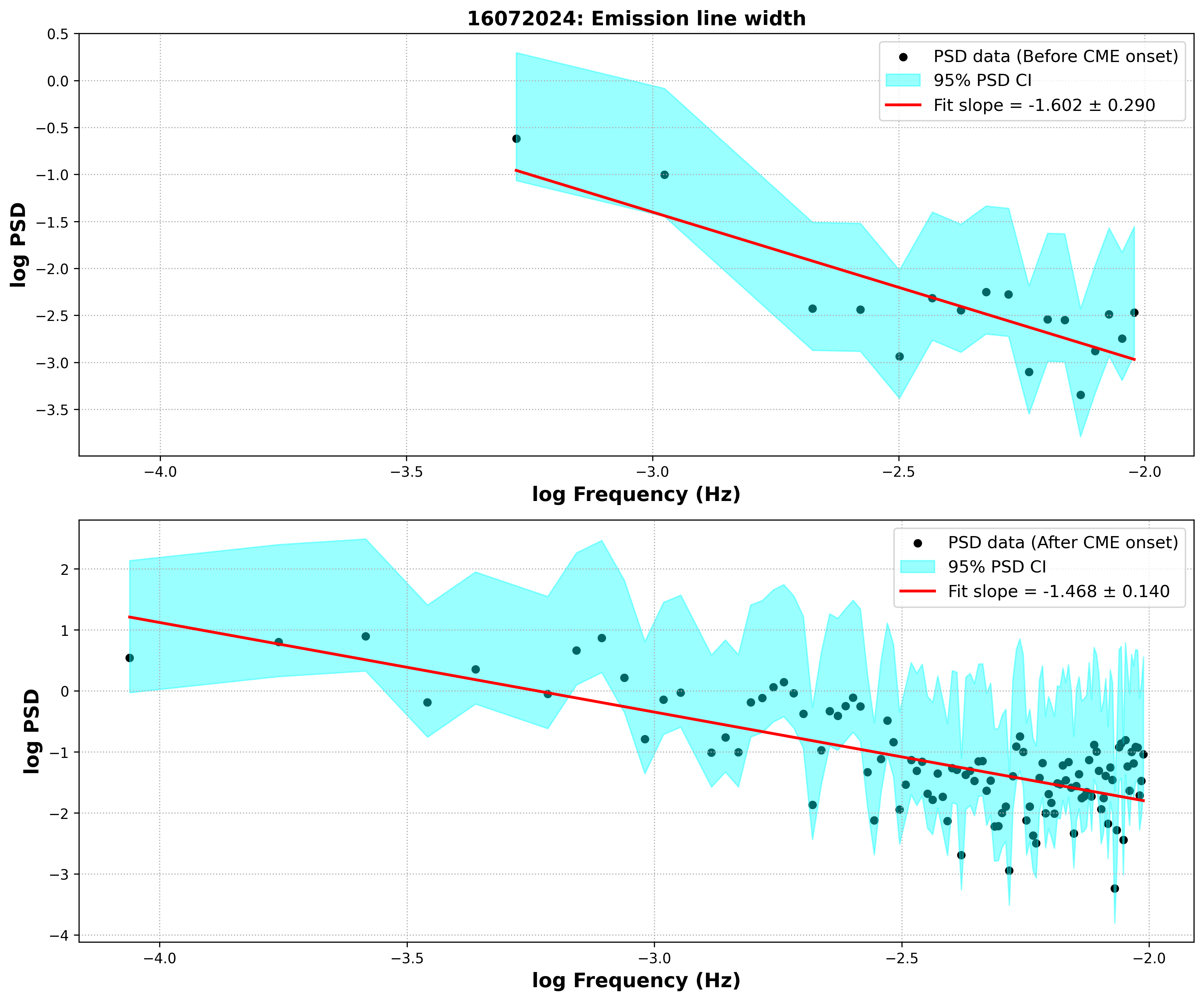}      
     \caption{Upper panel shows the Power Spectral Density (PSD) of the line width data for 16 July 2024 VELC 5303{\AA} observations, prior to 13:18\,UT (i.e., before the CME onset, red colour profile in the middle panel of Figure \ref{fig:light_curve_16072024}). Lower panel shows the PSD of the data after 
     13:18\,UT (i.e., post CME onset, blue colour profile in the middle panel of Figure \ref{fig:light_curve_16072024}). The red dashed lines in the two panels show the power law fit to the PSD data points, and the shaded area in light blue colour in the two panels indicate the 95\% confidence interval (CI).}
         \label{fig:psd}
\end{figure*}

We measured the slope of the PSD of the line widths (Figure \ref{fig:light_curve_16072024}), both before and after the coronal dimming. The results are shown in Figure \ref{fig:psd}. Note that the PSD analysis was carried out after removing the 13 \& 33\,min  periodicities mentioned earlier. 
The steps involved in the removal are Fourier transformation of the original time series data, filtering the frequency components corresponding to the time domain periodicities, and inverse Fourier transformation of the filtered data back to the time domain.

We filtered frequency components corresponding to the time bins 13${\pm}$2\,min \& 33${\pm}$2\,min since the above two periodicities show a spread in their corresponding frequencies. The upper panel shows the PSD before 13:18 UT (i.e., prior to the CME onset), while the lower panel shows the PSD after the CME onset. In both panels, the dots represent the data points and the red dashed lines indicate their power law fit.
Their slopes are 
$-1.60{\pm}0.29$ and $-1.47{\pm}0.14$, 
before and after the CME onset, respectively. 
Recent statistical studies indicate that turbulence in the interplanetary CMEs are of Kolmogorov type, and the spectral index in the inertial range varies between -1.38 and -1.82 \citep{Shaikh2024}. The origin of these fluctuations can be traced back
to the turbulent motions in the solar corona, from where the
solar wind is believed to be generated \citep{Yadav2025}. The slopes measured by the latter are in the range -1.47 to -1.83. Our estimates are consistent with the above results. 

This indicates that the estimated line widths, broader than the typical thermal widths ${\approx}0.68{\AA}$ for the formation temperature of 1.8 MK \citep{Cont2004,Kim2000, Kou2019}, are mostly due to turbulence. It is known that excess line broadening (nonthermal width) can occur due to turbulence \citep{Doschek2007}. Turbulence and magnetic reconnection are closely associated \citep{Lazarian1999}. Magnetic reconnections are driven by turbulence \citep{Xie2024}. These are consistent with the earlier reports that the unresolved nonthermal motions are suited for Kolmogorov type turbulence \citep{Chae1998}.

In the present case the line widths show more broadening after the CME. Based on the approximately same Kolmogorov slope of the PSD before and after the CME mentioned in the previous paragraph, we interpret that the increased line broadening in the post-CME phase is due to enhanced turbulence associated with coronal dimming and subsequent coronal magnetic field reconfiguration.
%, reconnections. 
Enhanced turbulence introduces a wider range of unresolved nonthermal velocity components along the line of sight. The post-CME observations of type I radio noise storm continum near the location of the CMEs is in support of the reconfiguration scenario mentioned above \citep{Ramesh1999,Kathiravan2007}. 

\begin{figure*}[!ht]
	\centering
	\includegraphics[width=0.6\linewidth]
	%{Dual_PSD_33_13_Removed_FWHM_05082024.png}
	%{05082024_Dual_PSD_33_13_Cleaned_width.png}
	{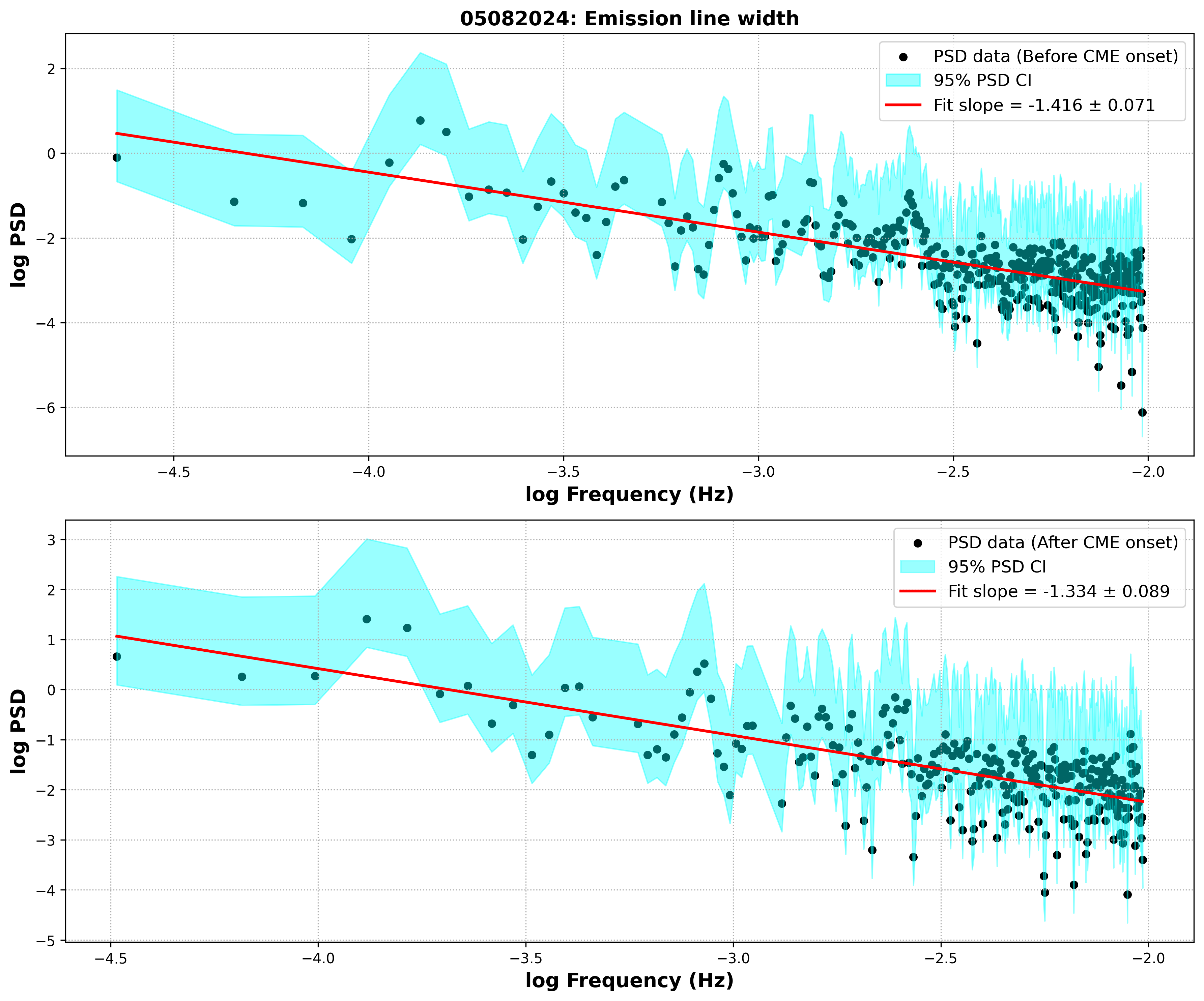}
	\caption{Upper panel shows the PSD of the line width data for 05 August 2024 VELC 5303{\AA} observations prior to 13:28\,UT (i.e., before CME onset, red colour profile in the  middle panel of Figure \ref{fig:light_curve_05082024}). Lower panel shows the PSD of the data post 13:28\,UT (i.e., after CME onset, blue colour profile in the middle panel of Figure \ref{fig:light_curve_05082024}). The red dashed lines in the two panels show the power law fit to the PSD data points, and the shaded area in light blue colour in the two panels indicate the 95\% CI. }
	\label{fig:psd1}
\end{figure*}

To confirm the above result, we studied another CME dimming event observed on 5 August 2024. On that day, the CME onset occurred at ${\approx}$13:28 UT. As shown in Figure \ref{fig:light_curve_05082024}, coronal dimming is clearly evident following the CME onset. We measured the PSD slopes before and after the CME, after excluding the 13 \& 33\,min periodicities from the data set similar to the 16 July 2024 event. Before the CME onset at 13:28 UT, the PSD slope is 
$-1.42{\pm}0.07$ (upper panel in Figure \ref{fig:psd1}). After the CME, the slope
is $-1.33{\pm}0.09$ (lower panel in Figure \ref{fig:psd1}). These PSD slopes
before and after the CME are reasonably consistent with our interpretations for the 16 July 2024 event mentioned in the previous paragraph. The shallower slopes compared to the 16 July 2024 event are possibly due to noise in the data (Figures \ref{fig:light_curve_16072024} \& \ref{fig:light_curve_05082024}). 

\begin{figure*}[!ht]
	\centering
	\includegraphics[width=0.6\linewidth]
	{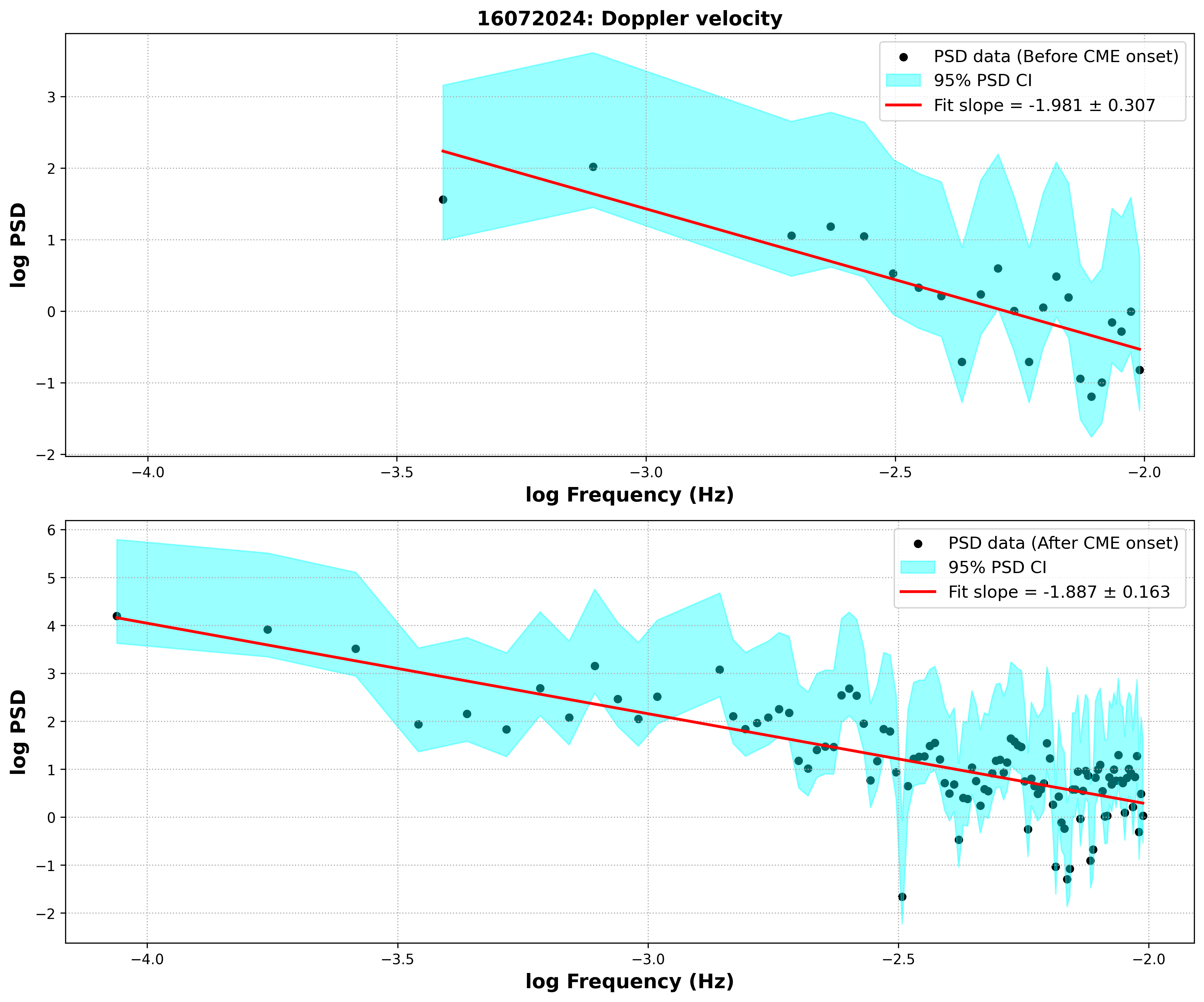}
%	{16072024_Dual_PSD_Cleaned_Doppler.png}
%	{Dual_PSD_33_13_Removed_doppler_16072024.png}
		\caption{Upper panel shows the PSD of the Doppler velocity data for 16 July 2024 VELC 5303{\AA} observations prior to 13:28\,UT (i.e., before CME onset, red colour profile in the lower panel of Figure \ref{fig:light_curve_16072024}). Lower panel shows the PSD of the data after 13:28\,UT (i.e., post CME onset, blue colour profile in the lower panel of Figure \ref{fig:light_curve_16072024}). The red dashed lines in the two panels show the power law fit to the PSD data points, and the shaded area in light blue colour in the two panels indicate the 95\% CI.} 
	\label{fig:psd2}
\end{figure*}

The PSD of line widths represents the
amplitude of line width fluctuations at various spatial/temporal scales. In a turbulent
system the energy is transferred from large length scales to
small-scale structures, i.e., a turbulent energy cascade, and eventually dissipates into heat energy. The largest structures or the energy-containing eddies injects energy into a broadband scale-to-scale transfer process.
So, we considered the PSD of the Doppler velocities also since it relates to the energy of turbulent eddies. For the 16 July 2024 event, we found that the slopes of before and after the CME onset are 
$-1.98{\pm}0.31$ and $-1.89{\pm}0.16$,
%$-2.16{\pm}0.36$ and $-1.89{\pm}0.17$,
%$–2.0 \pm 0.36$ and $–2.04 \pm 0.17$, 
respectively (Figure \ref{fig:psd2}). There is consistency between the values in the pre-CME and post-CME onset phases, similar to that of the line widths mentioned earlier. The fluctuations in the Doppler velocities are slightly steeper compared to that of the Doppler widths, but are in the reported range of slopes (-5/3 to -2) for solar wind velocity fluctuations \citep{Bavassano2005}.
We carried out PSD calculations for the Doppler velocities in the 5 August 2024 event also. It shows a nearly flat spectrum, probably due to the smaller change in the corresponding velocity values mentioned earlier. We intend to extend the study with larger data set to verify the above results.

\section{Summary and conclusions}
The Visible Emission Line Coronagraph (VELC) onboard Aditya-L1 is the first space solar coronagraph developed at the Indian Institute of Astrophysics. VELC has demonstrated its capability to observe emission line peak intensity, line widths, and Doppler velocities using the 5303{\AA} spectral line in the FoV spanning 1.05\,-\,1.5$R_{\odot}$. Recently, \citet{Ramesh2024} reported a CME associated coronal dimming event observed on 16 July 2024. The authors noted that the line widths increased by approximately 15\% after the CME. It was suggested that this enhancement in line width might be due to increased turbulence in the inner solar corona. In this article, we use sit \& stare mode observations carried out on 16 July 2024 and 5 August 2024 to examine the spectral line widths both before and after CMEs. The power spectral density (PSD) derived from these line widths exhibits a power-law distribution. Interestingly, the PSD slopes remained consistent before and after the CME events, closely matching the Kolmogorov slope of –5/3. However considering that the measured mean slopes vary in the 
range $-$1.33 to $-$1.98, it is possible that Iroshnikov-Kraichnan turbulence (slope\,${\approx}$\,$-$3/2; \citealp{Iroshnikov1964,Kraichnan1965}) and shock turbulence (slope\,${\approx}$\,$-$2; see, e.g. \citealp{Huang2020}) also play a role. Either way, these indicate that the observed line broadening is chiefly due to turbulence.
% as shown in Table \ref{tab:slopes}. 
Increase in emission line widths ${\approx}15\%$ (on 16 July 2024) and 
${\approx}7\%$ (on 5 August 2024) following the onset of the CME indicates enhanced turbulence due to the CME associated coronal dimming and subsequent coronal magnetic field reconfiguration.

\section*{Acknowledgements}
We thank P. Savarimuthu, S. Nagashree and and E. Yuvashree for their
assistance in processing the VELC data.
The VELC team members who developed the payload are thanked for their
efforts. Aditya-L1 is an observatory class mission, funded and operated by the Indian Space Research Organization (ISRO). Data obtained with the different payloads on board Aditya-L1 are archived at the Indian Space Science Data Centre (ISSDC) of ISRO. We thank the referees for their insightful comments 
to improve the manuscript.
\vspace{-1em}
\bibliography{main}{}

\begin{thebibliography}{}
\expandafter\ifx\csname natexlab\endcsname\relax\def\natexlab#1{#1}\fi

\bibitem[{{Bastian}(1994)}]{Bas94}
{Bastian}, T.~S. 1994, The Astrophysical Journal, 426, 774

\bibitem[{{Bavassano} {$et~al$.}(2005){Bavassano}, {Bruno}, \&
  {D'Amicis}}]{Bavassano2005}
{Bavassano}, B., {Bruno}, R., \& {D'Amicis}, R. 2005, Ann. Geophys., 23, 1025

\bibitem[{{Bemporad}(2008)}]{Bemporad2008}
{Bemporad}, A. 2008, \apj, 689, 572

\bibitem[{{Carley} {$et~al$.}(2021){Carley}, {Cecconi}, {Reid}, {Briand},
  {Sasikumar Raja}, {Masson}, {Dorovskyy}, {Tiburzi}, {Vilmer}, {Zucca},
  {Zarka}, {Tagger}, {Grie{\ss}meier}, {Corbel}, {Theureau}, {Loh}, \&
  {Girard}}]{Carley2021}
{Carley}, E.~P., {Cecconi}, B., {Reid}, H.~A., {$et~al$.} 2021, \apj, 921, 3

\bibitem[{{Chae} {$et~al$.}(1998){Chae}, {Sch{\" u}hle}, \&
  {Lemaire}}]{Chae1998}
{Chae}, J., {Sch{\" u}hle}, U., \& {Lemaire}, P. 1998, \apj, 505, 957

\bibitem[{{Contesse} {$et~al$.}(2004){Contesse}, {Koutchmy}, \&
  {Viladrich}}]{Cont2004}
{Contesse}, L., {Koutchmy}, S., \& {Viladrich}, C. 2004, Ann. Geophys., 22,
  3055

\bibitem[{{De Moortel} {$et~al$.}(2014){De Moortel}, {McIntosh}, {Threlfall},
  {Bethge}, \& {Liu}}]{Moortel2014}
{De Moortel}, I., {McIntosh}, S.~W., {Threlfall}, J., {Bethge}, C., \& {Liu},
  J. 2014, \apj, 782, L34

\bibitem[{{Doschek} {$et~al$.}(2007){Doschek}, {Mariska}, {Warren}, {Brown},
  {Culhane}, {Hara}, {Watanabe}, {Young}, \& {Mason}}]{Doschek2007}
{Doschek}, G.~A., {Mariska}, J.~T., {Warren}, H.~P., {$et~al$.} 2007, \apj,
  667, L109

\bibitem[{{Ghuge} {$et~al$.}(2025){Ghuge}, {Bhattacharjee}, \&
  {Subramanian}}]{Deep2025}
{Ghuge}, D., {Bhattacharjee}, D., \& {Subramanian}, P. 2025, \solphys, 300, 47

\bibitem[{{Huang} {$et~al$.}(2020){Huang}, {Wang}, {Sahraoui}, {Yuan}, {Liu},
  {Deng}, {Sun}, {Jiang}, {Xu}, {Yu}, {Wei}, \& {Zhang}}]{Huang2020}
{Huang}, S.~Y., {Wang}, Q.~Y., {Sahraoui}, Y., {$et~al$.} 2020, \apj, 891, 159

\bibitem[{{Ingale} {$et~al$.}(2015){Ingale}, {Subramanian}, \&
  {Cairns}}]{Ing2015a}
{Ingale}, M., {Subramanian}, P., \& {Cairns}, I. 2015, \mnras, 447, 3486

\bibitem[{{Iroshnikov}(1964)}]{Iroshnikov1964}
{Iroshnikov}, P.~S. 1964, Sov. Astron., 7, 566

\bibitem[{{Kathiravan} {$et~al$.}(2007){Kathiravan}, {Ramesh}, \&
  {Nataraj}}]{Kathiravan2007}
{Kathiravan}, C., {Ramesh}, R., \& {Nataraj}, H.~S. 2007, \apj, 656, L37

\bibitem[{Kim(2000)}]{Kim2000}
Kim, I. 2000, in Astron. Soc. Pacific Conf. Ser., Vol. 205, The Last Total
  Solar Eclipse of the Millennium in Turkey, ed. W.~W.~Livingston \& A.~{\"
  O}zgu{\" c}, 69

\bibitem[{Kolmogorov(1941)}]{kolmogorov1941}
Kolmogorov, A.~N. 1941, Doklady Akademiia Nauk SSSR, 30, 301

\bibitem[{{Koutchmy} {$et~al$.}(2019){Koutchmy}, {Baudin}, {Abdi}, {Golub}, \&
  {S{\`e}vre}}]{Kou2019}
{Koutchmy}, S., {Baudin}, F., {Abdi}, S., {Golub}, L., \& {S{\`e}vre}, F. 2019,
  \aap, 632, A86

\bibitem[{{Kraichnan}(1965)}]{Kraichnan1965}
{Kraichnan}, R.~H. 1965, Phys. Fluids, 8, 1385

\bibitem[{{Lazarian} \& {Vishniac}(1999)}]{Lazarian1999}
{Lazarian}, C., \& {Vishniac}, E.~T. 1999, \apj, 517, 700

\bibitem[{{Manoharan} {$et~al$.}(2000){Manoharan}, {Kojima}, {Gopalswamy},
  {Kondo}, \& {Smith}}]{Man2000}
{Manoharan}, P.~K., {Kojima}, M., {Gopalswamy}, N., {Kondo}, T., \& {Smith}, Z.
  2000, \apj, 530, 1061

\bibitem[{{Mishra} {$et~al$.}(2024){Mishra}, {Sasikumar Raja}, {Sanal
  Krishnan}, {Suresh Narra}, {Bhavana Hegde}, {Utkarsha}, {Muthu Priyal},
  {Pawan Kumar}, {Natarajan}, {Raghavendra Prasad}, {Singh}, {Umesh Kamath},
  {Kathiravan}, {Vishnu}, {Savarimuthu}, {Desai}, {Kumaran}, {Sagar}, {Kumar},
  {Bamrah}, \& {Kumar}}]{Mishra2024}
{Mishra}, S., {Sasikumar Raja}, K., {Sanal Krishnan}, V.~U., {$et~al$.} 2024,
  Exp. Astron., 57, 7

\bibitem[{{Muthu Priyal} {$et~al$.}(2025){Muthu Priyal}, {Ramesh}, {Singh}, \&
  {Sasikumar Raja}}]{MP2025ApJ...983..171M}
{Muthu Priyal}, V., {Ramesh}, R., {Singh}, J., \& {Sasikumar Raja}, K. 2025,
  \apj, 983, 171

\bibitem[{{Parate} {$et~al$.}(2025){Parate}, {Shaji}, {Srikanth},
  {Sankarasubramanian}, {Patnaik}, \& {Narendra}}]{Parate2025}
{Parate}, N.~P., {Shaji}, N., {Srikanth}, M., {$et~al$.} 2025, \solphys, 300,
  128

\bibitem[{{Perez} {$et~al$.}(2021){Perez}, {Bourouaine}, {Chen}, \&
  {Raouafi}}]{Perez2021}
{Perez}, J.~C., {Bourouaine}, S., {Chen}, C. H.~K., \& {Raouafi}, N.~E. 2021,
  \aap, 650, A22

\bibitem[{{Raghavendra Prasad} {$et~al$.}(2017){Raghavendra Prasad},
  {Banerjee}, {Singh}, {Nagabhushana}, {Kumar}, {Kamath}, {Kathiravan},
  {Venkata}, {Rajkumar}, {Natarajan}, {Juneja}, {Somu}, {Pant}, {Shaji},
  {Sankarsubramanian}, {Patra}, {Venkateswaran}, {Adoni}, {Narendra},
  {Haridas}, {Mathew}, {Mohan Krishna}, {Amareswari}, \& {Jaiswal}}]{BRP2017}
{Raghavendra Prasad}, B., {Banerjee}, D., {Singh}, J., {$et~al$.} 2017, Curr.
  Sci., 113, 613

\bibitem[{{Raghavendra Prasad} {$et~al$.}(2023){Raghavendra Prasad}, {Suresh},
  {Natarajan}, {Pawan Kumar}, {Kamath}, {Mishra}, {Bhavana Hegde}., {Sasikumar
  Raja}, \& {Singh}}]{BRP2023}
{Raghavendra Prasad}, B., {Suresh}, N.~V., {Natarajan}, V., {$et~al$.} 2023, J.
  Astron. Tele. Instr. Sys., 9, 044001

\bibitem[{{Raj Kumar} {$et~al$.}(2018){Raj Kumar}, {Prasad}, {Singh}, \&
  {Suresh}}]{RK2018}
{Raj Kumar}, N., {Prasad}, B.~R., {Singh}, J., \& {Suresh}, N.~V. 2018, Exp.
  Astron., 45, 219

\bibitem[{{Ramesh} {$et~al$.}(2023){Ramesh}, {Kathiravan}, \&
  {Kumari}}]{Ramesh2023}
{Ramesh}, R., {Kathiravan}, C., \& {Kumari}, A. 2023, \apj, 943, 43

\bibitem[{{Ramesh} {$et~al$.}(2001){Ramesh}, {Kathiravan}, \&
  {Sastry}}]{Ram2001}
{Ramesh}, R., {Kathiravan}, C., \& {Sastry}, {\relax Ch}.~V. 2001, \apj, 548,
  L229

\bibitem[{{Ramesh} {$et~al$.}(2024){Ramesh}, {Muthu Priyal}, {Singh},
  {Sasikumar Raja}, {Savarimuthu}, \& {Gavshinde}}]{Ramesh2024}
{Ramesh}, R., {Muthu Priyal}, V., {Singh}, J., {$et~al$.} 2024, \apjl, 976, L6

\bibitem[{{Ramesh} {$et~al$.}(1999){Ramesh}, {Subramanian}, \&
  {Sastry}}]{Ramesh1999}
{Ramesh}, R., {Subramanian}, K.~R., \& {Sastry}, {\relax Ch}.~V. 1999,
  \solphys, 185, 77

\bibitem[{{Sasikumar Raja} {$et~al$.}(2016){Sasikumar Raja}, {Ingale},
  {Ramesh}, {Subramanian}, {Manoharan}, \& {Janardhan}}]{KSR2016}
{Sasikumar Raja}, K., {Ingale}, M., {Ramesh}, R., {$et~al$.} 2016, J. Geophys.
  Res. (Space Phys.), 121, 11605

\bibitem[{{Sasikumar Raja} {$et~al$.}(2019){Sasikumar Raja}, {Subramanian},
  {Ingale}, \& {Ramesh}}]{KSR2019}
{Sasikumar Raja}, K., {Subramanian}, P., {Ingale}, M., \& {Ramesh}, R. 2019,
  \apj, 872, 77

\bibitem[{{Sasikumar Raja} {$et~al$.}(2021){Sasikumar Raja}, {Subramanian},
  {Ingale}, {Ramesh}, \& {Maksimovic}}]{KSR2021}
{Sasikumar Raja}, K., {Subramanian}, P., {Ingale}, M., {Ramesh}, R., \&
  {Maksimovic}, M. 2021, \apj, 914, 137

\bibitem[{{Sasikumar Raja} {$et~al$.}(2017){Sasikumar Raja}, {Subramanian},
  {Ramesh}, {Vourlidas}, \& {Ingale}}]{KSR2017}
{Sasikumar Raja}, K., {Subramanian}, P., {Ramesh}, R., {Vourlidas}, A., \&
  {Ingale}, M. 2017, \apj, 850, 129

\bibitem[{{Sasikumar Raja} {$et~al$.}(2022){Sasikumar Raja}, {Suresh}, {Singh},
  \& {Prasad}}]{KSR2022}
{Sasikumar Raja}, K., {Suresh}, N.~V., {Singh}, J., \& {Prasad}, B.~R. 2022,
  Adv. Space Res., 69, 814

\bibitem[{{Shaikh}(2024)}]{Shaikh2024}
{Shaikh}, Z.~I. 2024, \mnras, 530, 3005

\bibitem[{{Sharma} \& {Morton}(2023)}]{Sharma2023}
{Sharma}, R., \& {Morton}, R.~J. 2023, Nat. Astron., 7, 1301

\bibitem[{{Shen} {$et~al$.}(2023){Shen}, {Polito}, {Reeves}, {Chen}, {Yu}, \&
  {Xie}}]{Shen2023}
{Shen}, C., {Polito}, V., {Reeves}, K.~K., {$et~al$.} 2023, Front. Astron.
  Space Sci., 10, 19

\bibitem[{{Singh} {$et~al$.}(2013){Singh}, {Bayanna}, \&
  {Sankarasubramanian}}]{JS2013}
{Singh}, J., {Bayanna}, A., \& {Sankarasubramanian}, K. 2013, J. Opt., 42, 5

\bibitem[{{Singh} {$et~al$.}(2025){Singh}, {Ramesh}, {Prasad}, {Muthu Priyal},
  {Sasikumar Raja}, \& {Suresh Venkata}}]{Singh2025}
{Singh}, J., {Ramesh}, R., {Prasad}, B.~R., {$et~al$.} 2025, Solar Phys., 300,
  66

\bibitem[{{Singh} {$et~al$.}(2006){Singh}, {Sakurai}, \&
  {Ichimoto}}]{Singh2006}
{Singh}, J., {Sakurai}, T., \& {Ichimoto}, K. 2006, \apj, 51, 639

\bibitem[{{Singh} {$et~al$.}(2004){Singh}, {Sakurai}, {Ichimoto}, \&
  {Watanabe}}]{Singh2004}
{Singh}, J., {Sakurai}, T., {Ichimoto}, K., \& {Watanabe}, T. 2004, \apj, 617,
  L81

\bibitem[{{Suresh Venkata} {$et~al$.}(2023){Suresh Venkata}, {Bhavana Hegde}.,
  {Utkarsha}, {Natarajan}, {Pawan Kumar}, {Kamath}, \& {Prasad}}]{VSN2023a}
{Suresh Venkata}, N., {Bhavana Hegde}., {Utkarsha}, D., {$et~al$.} 2023, J.
  Astron. Instr., 12, 2350011

\bibitem[{{Suresh Venkata} {$et~al$.}(2022){Suresh Venkata}, {Prasad}, \&
  {Singh}}]{VSN2022a}
{Suresh Venkata}, N., {Prasad}, B.~R., \& {Singh}, J. 2022, Exp. Astron., 53,
  71

\bibitem[{{Suresh Venkata} {$et~al$.}(2024){Suresh Venkata}, {Sasikumar Raja},
  {Prasad}, {Singh}, {Mishra}, {Krishnan}, {Hegde}, {D.}, {V.}, {S.}, {V.},
  {P.}, {Gavshinde}, \& {Kamath}}]{VSN2024}
{Suresh Venkata}, N., {Sasikumar Raja}, K., {Prasad}, B.~R., {$et~al$.} 2024,
  Exp. Astron., 58, 6

\bibitem[{{Taylor}(1938)}]{Taylor1938}
{Taylor}, G.~I. 1938, Proc. Roy. Soc. Lon., Ser. A, Math. Phys. Sci., 164, 476

\bibitem[{{Vipin Yadav} {$et~al$.}(2025){Vipin Yadav}, {Choudhury}, \&
  {Goutham}}]{Yadav2025}
{Vipin Yadav}, K., {Choudhury}, R.~K., \& {Goutham}, M.~M. 2025, \apj, 991, 181

\bibitem[{{Xie} {$et~al$.}(2024){Xie}, {Li}, {Reeves}, \& {Gou}}]{Xie2024}
{Xie}, X., {Li}, G., {Reeves}, K.~K., \& {Gou}, T. 2024, Fron. Astron. Space
  Sci., 11, 1383746

\end{thebibliography}
\bibliographystyle{apj}
\end{document}